\documentclass[fleqn,12pt,twoside]{article}
\usepackage{espcrc1}

\usepackage{graphicx}
\usepackage[figuresright]{rotating}

\newcommand{\AmS}{{\protect\the\textfont2
  A\kern-.1667em\lower.5ex\hbox{M}\kern-.125emS}}

\title{Three-Body Elastic and Inelastic Scattering at Intermediate Energies}

\author{H. Liu\address{NSCL, Michigan State University, East Lansing, MI 48824, USA},
        Ch. Elster\address{Dept. of Physics \& Astronomy, Ohio University,  Athens, OH 45701, USA},
        and
        W. Gl\"ockle\address{ Theoretische Physik II, Ruhr-Universit\"at Bochum, D-44780 Bochum, Germany}
       } 
\begin{document}

\maketitle

\begin{abstract}
{\bf Abstract:} The Faddeev equation for three-body scattering at arbitrary energies is formulated in
momentum space and directly solved in terms of momentum vectors without employing a partial
wave decomposition. For identical bosons this results in a three-dimensional integral
equation in five variables, magnitudes of relative momenta and angles. The cross sections for
both elastic and breakup processes in the intermediate energy range up to about 1 GeV are
calculated based on a Malfliet-Tjon type potential, and the convergence of the multiple
scattering series is investigated. 
\end{abstract}

\section{Introduction}
Traditionally three-nucleon scattering calculations are carried out by
solving Faddeev equations in a partial wave truncated basis. That means an
angular momentum decomposition replaces the continuous angle variables by discrete
orbital angular momentum quantum numbers, and thus reduces the number
of continuous variables needed to be discretized in a numerical
treatment. For low projectile energies the procedure of considering
orbital angular momentum components appears physically justified due
to arguments related to the centrifugal barrier. 
If one considers three-nucleon scattering at a few hundred
MeV projectile energy, the number of partial waves needed to achieve
convergence proliferates, and limitations with respect to
computational feasibility and accuracy begin to appear. The amplitudes
acquire stronger angular dependence, which is already visible in the
two-nucleon amplitudes, and their formation by an increasing number of
partial waves not only becomes more tedious but also less
informative. 
It appears therefore natural to avoid a partial wave representation
completely and work directly with vector variables.

\section{Selected Three-Body Scattering Observables}

In its simplest form the Faddeev equation with scalar particles is a
three-dimensional integral equation in in five variables, which is numerically
solved below and above the break-up threshold. From its solution the scattering
amplitude is obtained as function of vector Jacobi momenta. As a simplification
we neglect spin and isospin degrees of freedom and treat three-boson scattering.
The interaction employed is of Malfliet-Tjon type, i.e. consists of a short range
repulsive and intermediate range attractive Yukawa force. The parameters of the
potential are adjusted so that a bound state at $E_d$=-2.23~MeV is supported.
The numerical feasibility and stability of our algorithm for solving the Faddeev
equations, especially the treatment of the logarithmic singularities with a spline
based semi-analytic method is demonstrated in Refs.~\cite{scatter3d,liuthesis}

The scattering
amplitude is then used to calculate either elastic scattering observables, total
and differential cross sections, as well as break-up observables, i.e. exclusive
and inclusive inelastic scattering cross sections. In Fig.~1 the elastic
differential cross section is shown as function of center-of-mass (c.m.)
scattering angle for energies between 0.2 and 1.0~GeV.

\vspace*{-4mm}
\begin{figure}[htb!]
  \includegraphics[width=55mm,angle=-90]{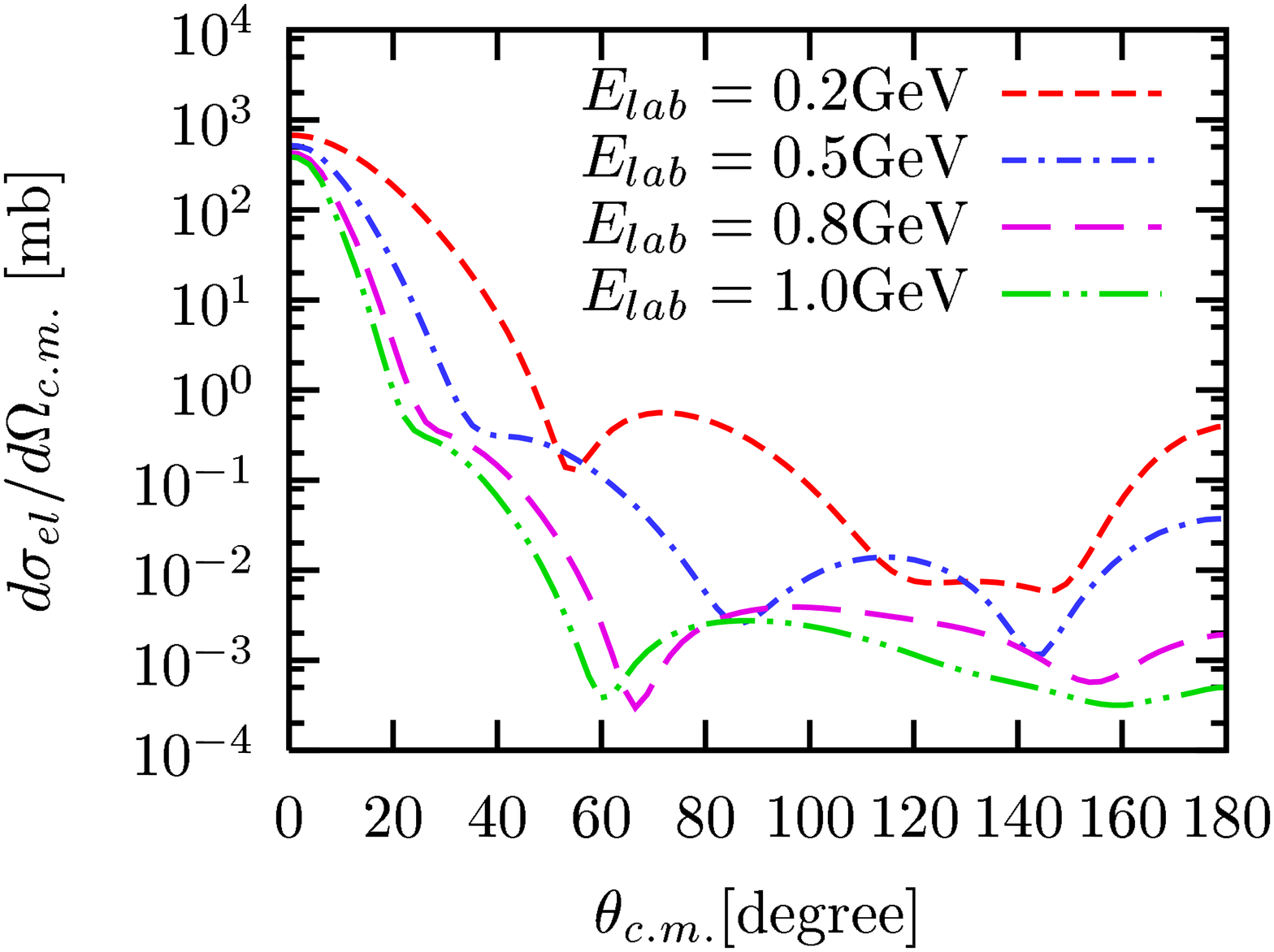}
\end{figure}
\label{fig1}
\vspace*{-6.0cm}\leftskip=7.5cm
\begin{minipage}[htb!]{7.0cm}
{{\bf Fig.~1: }\footnotesize
The elastic differential cross section (c.m.) at $E_{lab}=$~0.2~GeV,
0.5~GeV, 0.8~GeV, and 1.0~GeV projectile energy as function of the scattering
angle $\theta_{c.m.}$. All calculations are solutions of the full Faddeev
equation.}
\end{minipage}

\leftskip=0cm 
\vspace*{25mm}

The semi-exclusive cross section d(N,N') for scattering at 1~GeV is given in Fig.~2
for the emission angles 15$^o$ and 33$^o$, where the full Faddeev calculation together
with the lowest orders in the multiple scattering series are displayed. 

\leftskip=-0.5cm

\begin{minipage}[h]{8.7cm}
 \includegraphics[width=87mm,angle=-90]{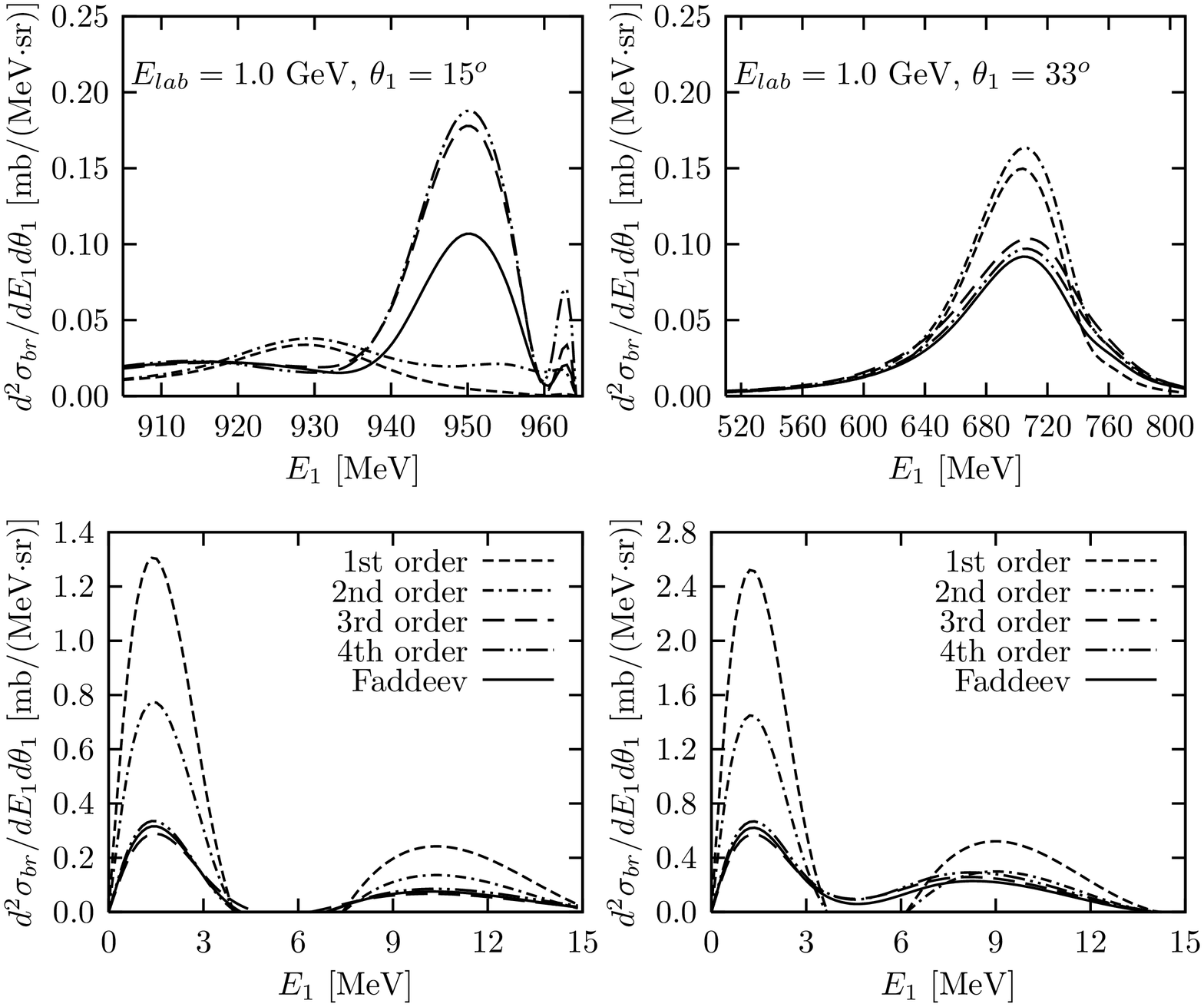}
\label{fig2}
\end{minipage}

\vspace*{-8.2cm}\leftskip=9.7cm
\begin{minipage}[h]{5.5cm}
{{\bf Fig.~2: }\footnotesize
The semi-exclusive cross section at 1.0~GeV laboratory incident energy and at 15$^o$ angle
(left panels) and 33$^o$ angle of the emitted particle (right panels). In both cases the
upper
panel displays the high energy range of the emitted particle, whereas the lower panel shows
the
low energy range. The full solution of the Faddeev equation is given by the solid line in all
panels, The contribution of the lowest orders of the
multiple scattering series added up successively is given by the other curves as indicated in
the legends.
}
\end{minipage}

\leftskip=0cm\vspace*{0.6cm}

The peak at the highest energy of the emitted particle is the so called final state
interaction (FSI) peak, which only develops if rescattering terms are taken into account.
This peak is a general feature of semi-exclusive scattering and is present at all energies. The
next peak is the so called quasi-free (QFS) peak, and one observes that at both angles one needs
at least rescattering up to the 3rd order to come close to the full result. At both angles the very
low energies of the emitted particle exhibit a strong peak in first order, which is considerably
lowered by the first rescattering. Here a calculation up to 3rd order in the multiple
scattering series is already sufficient. For  small ejectile angles there is interference
between the FSI and QFS peak resulting in a shift of the QFS peak to higher ejectile energies
when higher orders in the multiple scattering series are taken into account. This
phenomenon is not present once the angle of the ejected particle gets larger.

In an exclusive breakup process, two of the outgoing particles are measured in
coincidence, resulting in the five-fold differential break-up cross section.
As illustration  we give in Figs.~3 and 4  specific configurations at
$E_{lab} = 1.0$~GeV and consider the exclusive breakup cross section
$d^5 \sigma_{br} / d\Omega_p d\Omega_q dE_q$ given in the c.m. frame.
The energy of the outgoing particle is given by $E_q=\frac{3}{4m}q^2$ and takes
for $E_{lab}$=1.0~GeV the value $\frac{2}{3} E_{lab} + E_d \approx$~664~MeV.

\vspace*{-5mm}
\begin{figure}[htb!]
\includegraphics[width=80mm]{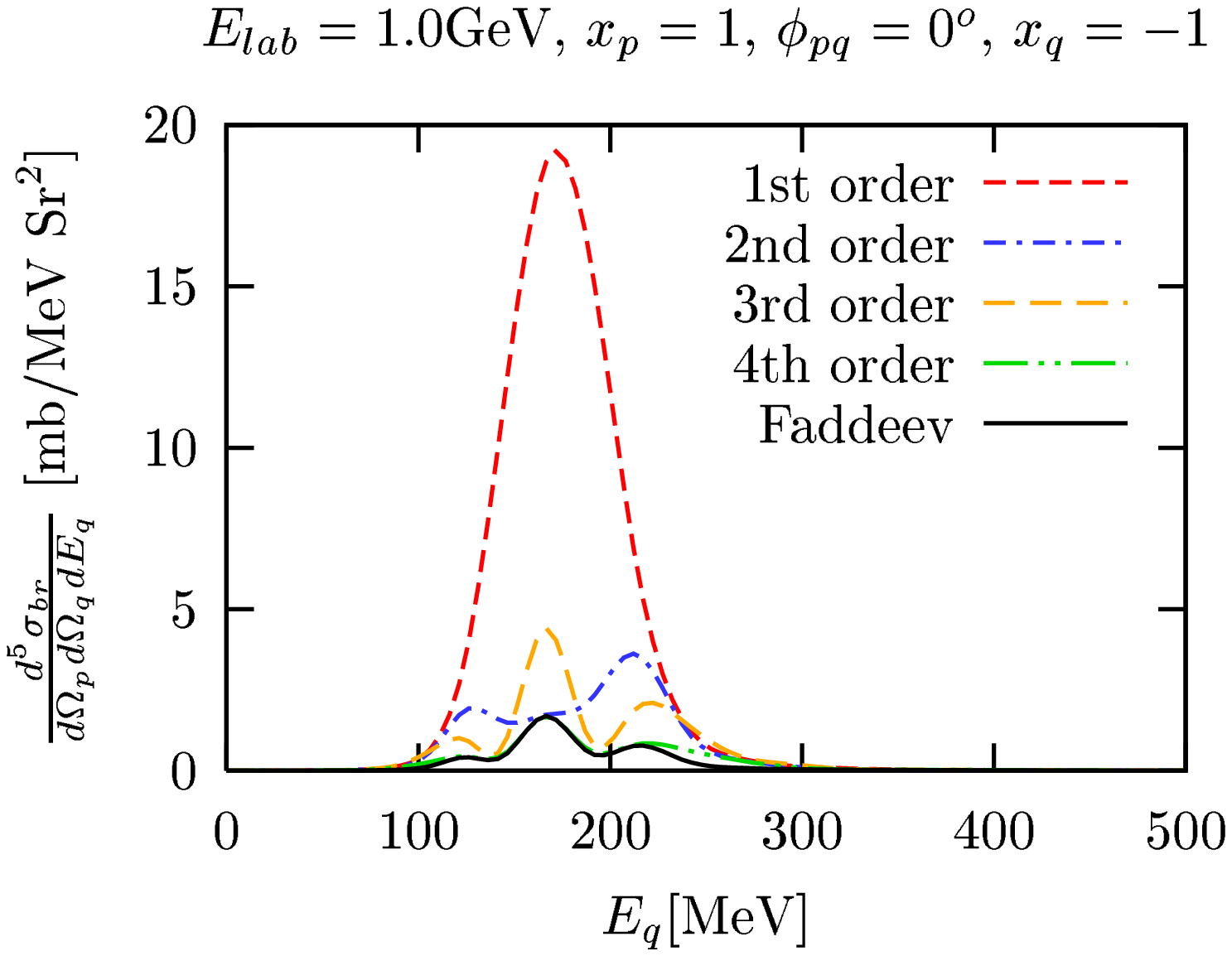}
\vspace*{-4.5cm} 
\includegraphics[width=80mm]{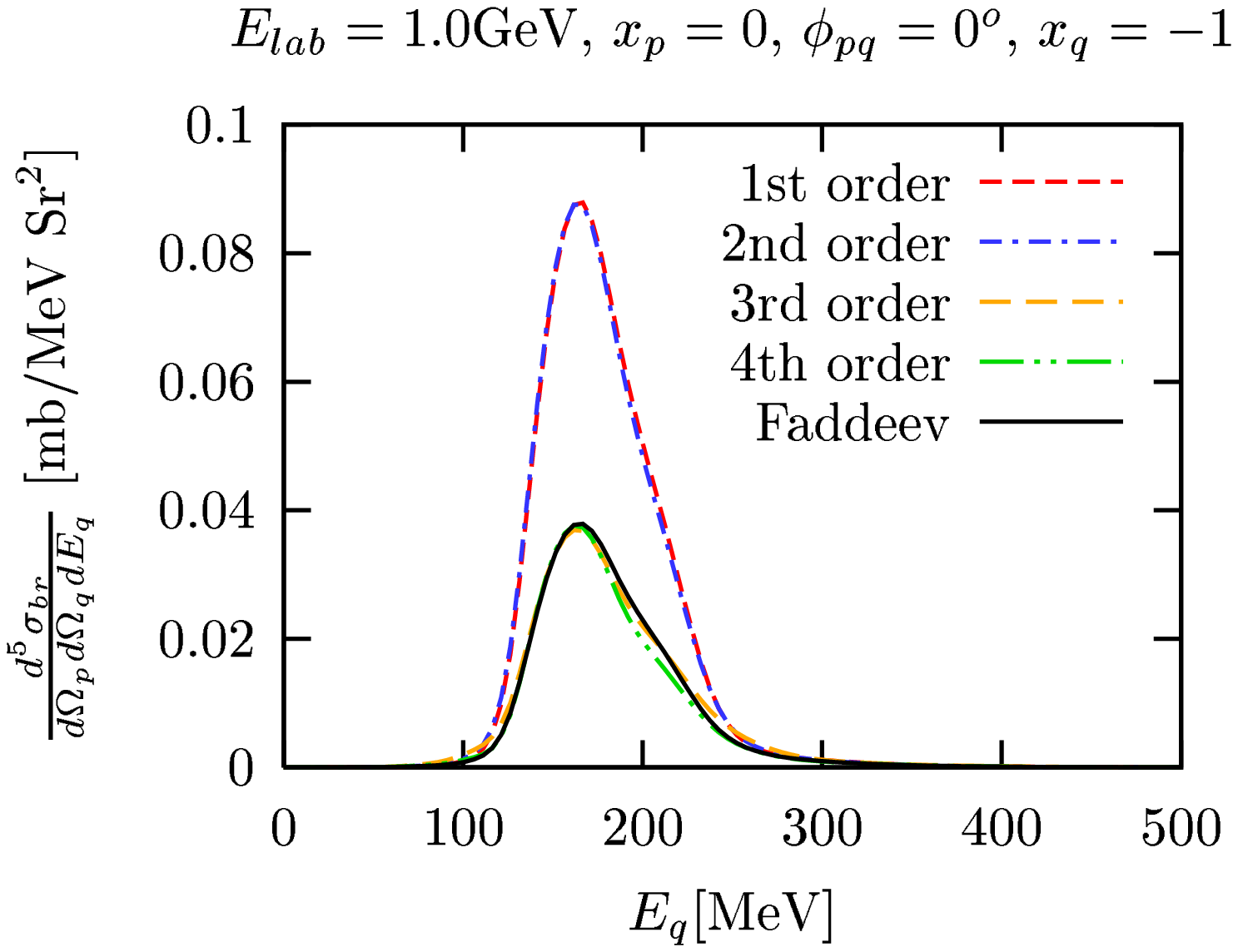}
\label{fig3}
\vspace*{4.0cm} \\
{{\bf Fig.~3:} \footnotesize
The exclusive differential cross section at $E_{lab} = 1.0$~GeV and
$x_q = -1, \phi_{pq} = 0^o$, and $x_p = 1$, indicating a collinear
condition (left panel). The right panel shows
the same but with $x_p = 0$, indicating that none of the outgoing particles is
collinear with the incoming one. The full solution of the Faddeev equation is given
by the solid line. The contributions of the lowest orders in the multiple
scattering series added up successively are given by the other curves as
indicated in the legend.}
\end{figure}

\leftskip=0cm\vspace*{-5mm}

In Fig.~3 we show two specific configurations illustrating quasi-free (QFS)
scattering conditions. The QFS condition assumes one particle at rest in the
laboratory frame, e.g. ${\bf k_1}=0$. This is equivalent to {\bf
q}=-$\frac{1}{2}{\bf q_0}$, which means that the cosine of the angle between
{\bf q} and {\bf q}$_0$, $x_q=-1$. The energy $E_q$ corresponding to the
QFS condition is approximately one quarter of the total energy, leading to
the peak at $\approx$~168~MeV. The left panel indicates very clearly that if one
of the scattered particle is parallel to the incoming beam, the first order is
dominant, whereas the rescattering terms significantly reduce this first order
peak. 

The situation is quite different if we consider socalled star configurations, where the three
outgoing particles have equal energies and leave with angles of 120$^o$ to each other in the
 c.m. frame. If the plane spanned by the three outgoing particles is orthogonal to
the beam direction, the configuration is named space star, if the beam lies in the plane it
is called coplanar star. These two special configurations are shown in Fig.~4.

\begin{figure}[htb!]
\includegraphics[width=80mm]{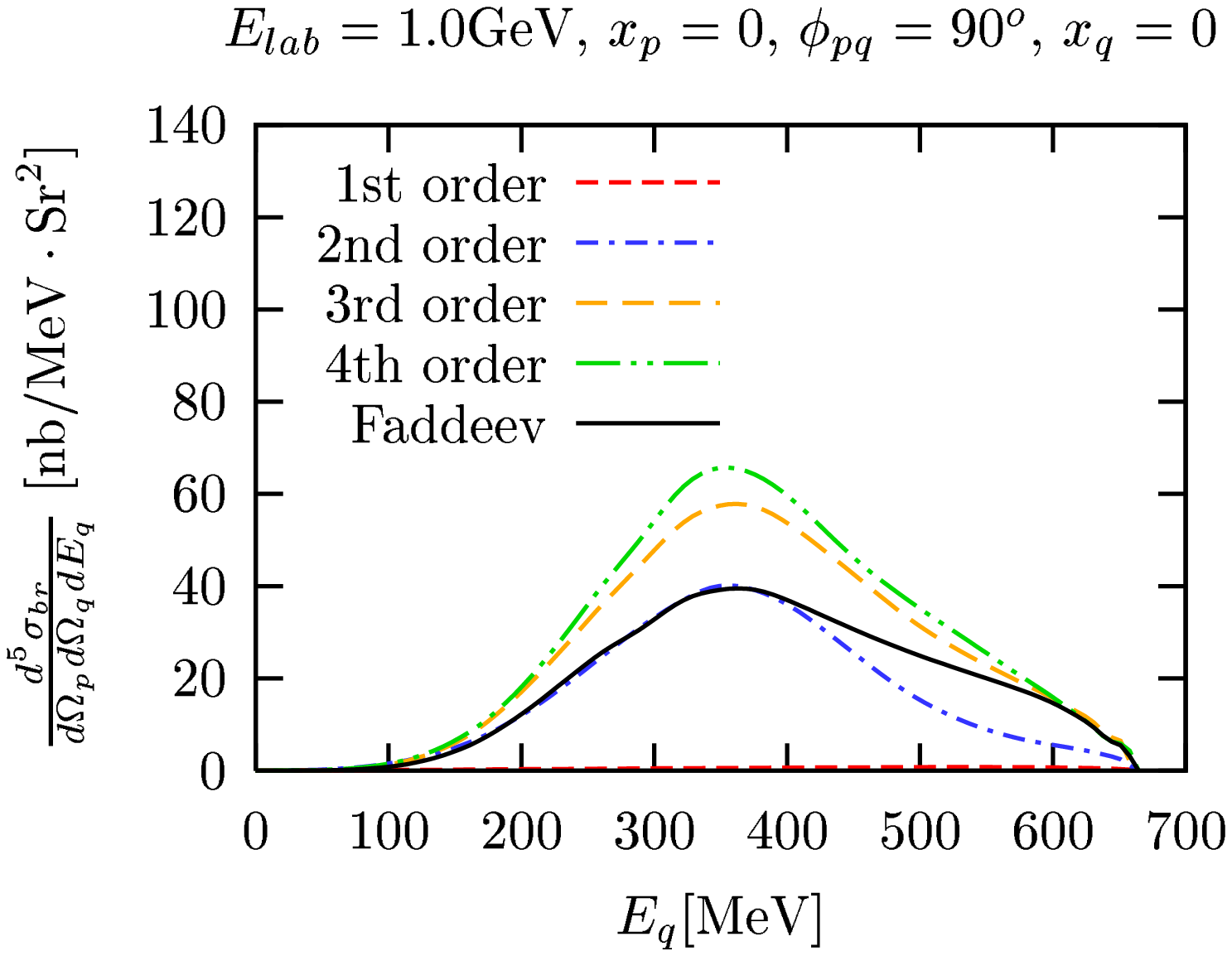}
\vspace*{-4.5cm} 
\includegraphics[width=80mm]{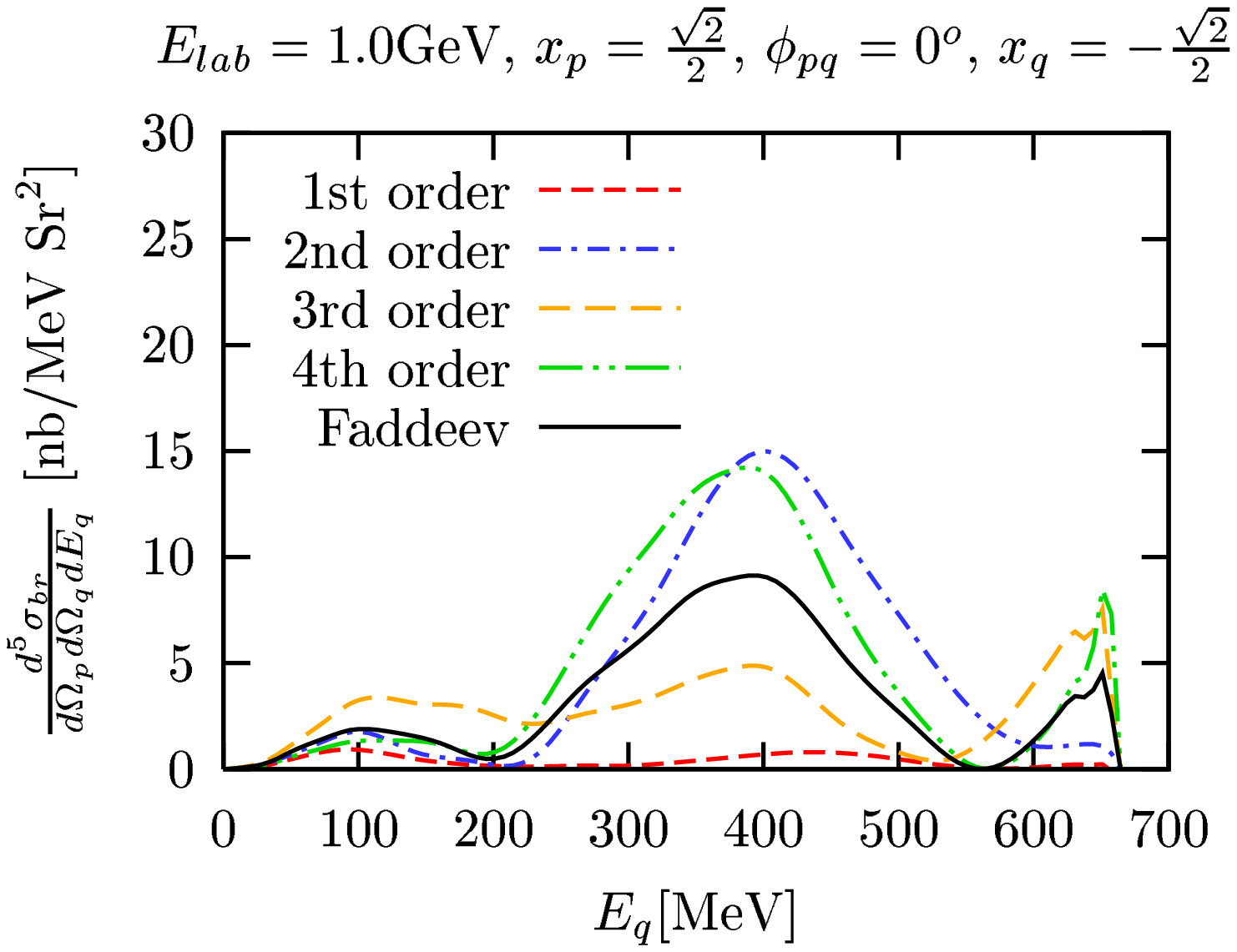}
\vspace*{4.0cm} \\
{{\bf Fig.~4:} \footnotesize
The exclusive differential cross section at $E_{lab} = 1.0$~GeV for the space star
(left panel) and the coplanar star configuration (right panel). The full solution of the
Faddeev equation is given
by the solid line. The contributions of the lowest orders in the multiple
scattering series added up successively are given by the other curves as
indicated in the legend.}
\label{fig4}
\end{figure}

\leftskip=0cm\vspace*{-9mm}

Since the momenta of the outgoing particles are equal for a given beam energy (here 1~GeV),
the energy of a single particle in the
star configuration is approximately $(\frac{2}{3} E_{lab} + E_d)/2$~=~332~MeV.
One important feature of the star configuration is clearly seen, the first order calculation
does essentially not contribute to the cross sections. The peak around 330~MeV is completely
developed by rescattering contributions. Fig.~4 also shows that for the space star the
rescattering contributions shown increase the cross section, whereas the full calculation is
lower. This indicates that the multiple scattering series converges very slowly.
The situation is similar for the coplanar star, where adding higher orders lets the cross
section oscillate around the final result. In the coplanar configuration FSI peaks 
develop at the highest and lowest energy $E_q$ when rescattering terms are taking into
account. 

The study of exclusive breakup processes shows very clearly that for the specific
configurations considered here, even at an
energy as high as 1~GeV the full solution of the Faddeev calculation is needed
to obtain a converged result. Further studies scanning the complete three-body
phase space are underway. This will be important in order to shed light on
previous theoretical analyses of p(d,ppn) reactions which relied on low order
reaction mechanisms. It will be also important, to investigate of there are
regions in phase space
where low order calculations are valid.

\section{Relativistic Effects in First Order Calculations}

As stated in the introduction, the key advantage of our three-dimensional formulation lies in
its applicability at higher energies. At these energies relativistic effects are expected to
become important, and their influence on the observables needs to be studied. In our
approach we want to identify relativistic effects within the framework of Poincar{\'e}
invariant quantum mechanics. Here Poincar{\'e} invariance is an exact symmetry
that is realized by a unitary representation of the Poincar{\'e} group
on a few-particle Hilbert space~\cite{keipol1}.  The equations we use have the same
operator form as the non-relativistic Faddeev equations, however the ingredients
are quite different.

\vspace*{-4mm}
\begin{figure}[htb!]
  \includegraphics[width=84mm,height=55mm]{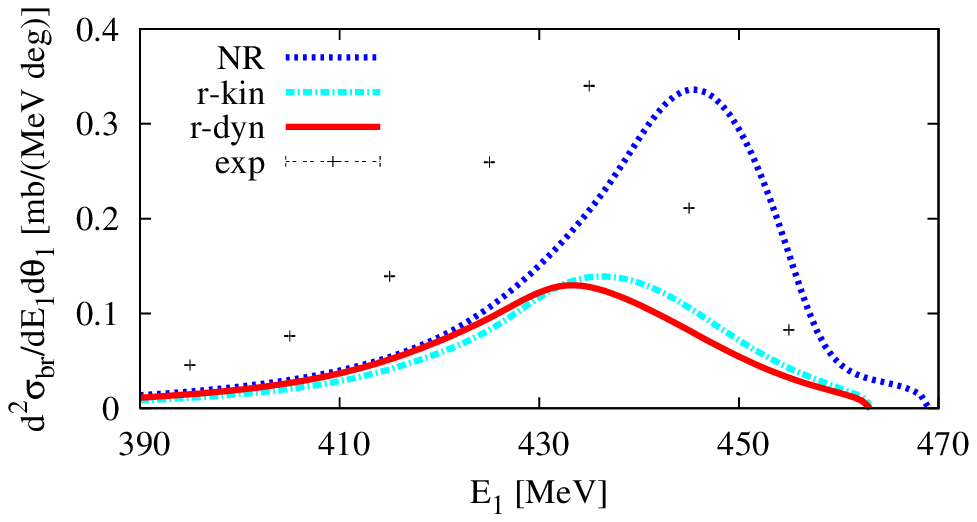}
\end{figure}
\label{fig5}
\vspace*{-6.5cm}\leftskip=8.5cm
\begin{minipage}[htb!]{6.7cm}
{{\bf Fig.~5: }\footnotesize
The semi-exclusive differential cross section at $E_{lab} = 495$~MeV and $\theta_{lab}=
18^o$ calculated with the first order term in the multiple scattering series. The dotted line
represents the non-relativistic calculation. Using relativistic kinematics leads the long
dashed line (r-kin), whereas a full relativistic calculation (r-dyn) gives the solid line. 
The data represent the spin averaged differential cross section for the (p,n) charge exchange
process \cite{chen}.}
\end{minipage}

\leftskip=0cm\vspace*{0.7cm}

First, there are the kinematic effects, which account for the Lorentz transformations between
the laboratory and c.m. frame of the three-body system, and different phase space factors
in the cross sections. Further Lorentz transformations occur when considering the definitions
of Jacobi momenta in the three-body system, which lead to a significantly 
more involved expression for the permutation operator for identical particles. 
Further relativistic effects arise from the propagators. Since in a first step we only
concentrate on a calculation based on the first order term in the multiple scattering
expansion, we only need to consider the two-body propagator in a relativistic
Lippmann-Schwinger (LS) equation. To compare with a non-relativistic calculation, this
relativistic LS equation needs a potential as driving term which is phase shift equivalent to
the non-relativistic one. 
Last, since in a relativistic formulation the two-body LS
equation depends on the two-body total momentum, it must be boosted. We follow here
the scheme described in Ref.~\cite{witala}. In Fig.~5 we present first results for the
semi-exclusive break-up cross section in comparison with a non-relativistic calculations.
Taking into account relativistic kinematics puts the position of the QFS peak consistent
with experimental information. Since in QFS kinematics one particle is assumed to be a
spectator, dynamic relativistic effects are expected to be small, as is confirmed by the
calculation. All calculations shown are in first order, and rescattering effects are still
important at this energy, no statement concerning the height of the peaks should be made.
However, we can see, that at higher energies relativistic effects will be quite visible,
and further investigation is under way.

\vspace*{3mm}
\begin{small} 
{\bf Acknowledgments:} We thank T. Lin for allowing to use her relativistic result for
comparison. This work was supported in part by the U.S. DOE under contract no.
DEFG02-93ER40576. We thank the Supercomputer Centers NERSC and OSC
 for the use of their facilities.
\end{small}


\begin{thebibliography}{9}
\bibitem{scatter3d}  H. Liu, Ch. Elster, W. Gl\"ockle, Phys. Rev. C{\bf 72}, 054003 (2005).
\bibitem{liuthesis} H. Liu, PhD Thesis, Ohio University, August 2005.
\bibitem{keipol1} See e.g. B.D. Keister and W.N. Polyzou, Adv. Nucl. Phys. {\bf 20}, 225,
1991.
\bibitem{chen} X.Y. Chen et {\it al.}, Phys. Rev. {\bf C47}, 2159 (1993).
\bibitem{witala} H. Witala, J. Golak, W. Gl\"ockle, H. Kamada, Phys. Rev. {\bf C71}, 054001
(2005).
\end{thebibliography}
\end{document}